\documentclass[11pt]{article}
\usepackage{epsfig}
\begin{document}

\def\wgta#1#2#3#4{\hbox{\rlap{\lower.35cm\hbox{$#1$}}
\hskip.2cm\rlap{\raise.25cm\hbox{$#2$}}
\rlap{\vrule width1.3cm height.4pt}
\hskip.55cm\rlap{\lower.6cm\hbox{\vrule width.4pt height1.2cm}}
\hskip.15cm
\rlap{\raise.25cm\hbox{$#3$}}\hskip.25cm\lower.35cm\hbox{$#4$}\hskip.6cm}}

\def\wgtb#1#2#3#4{\hbox{\rlap{\raise.25cm\hbox{$#2$}}
\hskip.2cm\rlap{\lower.35cm\hbox{$#1$}}
\rlap{\vrule width1.3cm height.4pt}
\hskip.55cm\rlap{\lower.6cm\hbox{\vrule width.4pt height1.2cm}}
\hskip.15cm
\rlap{\lower.35cm\hbox{$#4$}}\hskip.25cm\raise.25cm\hbox{$#3$}\hskip.6cm}}

\def\begeq{\begin{equation}}
\def\endeq{\end{equation}}
\def\zbar{\bar{z}}
\def\begeqar{\begin{eqnarray}}
\def\endeqar{\end{eqnarray}}
\def\partialbar{\bar{\partial}}
\def\wbar{\bar{w}}
\def\phibar{\bar{\phi}}
\def\xhat{\hat{x}}
\def\phat{\hat{p}}
\def\abar{\bar{a}}
\def\e{\epsilon}
\def\t{\theta}
\def\a{\alpha}
\def\k{\vec{k}}
\def\r{\vec{r}}

%
%
%
%

\title{The arboreal 
gas and  the 
supersphere sigma model}

\author{Jesper Lykke Jacobsen$^{a,b}$ and Hubert Saleur$^{b,c}$\\
\smallskip\\
$^{a}$LPTMS\\
B\^at.~100, Universit\'e Paris Sud\\
Orsay, 91405\\
France\\
\smallskip\\
$^{b}$ Service de Physique Th\'eorique\\
CEA Saclay\\
Gif Sur Yvette, 91191\\
France\\
\smallskip\\
$^{c}$ Department of Physics and Astronomy\\
University of Southern California\\
Los Angeles, CA 90089\\
USA\\
\smallskip\\
}

\maketitle

\begin{abstract}
We discuss  the relationship between the phase diagram of 
the $Q=0$ state Potts model, the arboreal gas model, and the 
supersphere sigma model $S^{0,2}=OSP(1/2)/OSP(0/2)$. We identify the 
Potts antiferromagnetic critical point with the critical point of the 
arboreal gas (at negative tree fugacity), and with a critical point of the 
sigma model. We 
show that the corresponding conformal theory on the square lattice 
has a non-linearly realized $OSP(2/2)=SL(1/2)$ symmetry, and involves 
non-compact degrees of freedom, with a continuous spectrum of 
critical exponents. The role of global topological properties in the 
sigma model transition is  discussed in terms of a generalized 
arboreal gas model. 

\end{abstract}

\section{Introduction}

The question of possible fixed points and RG flows in  $1+1$ theories with 
supergroup symmetries plays a major role
in the supersymmetric approach to phase transitions in non-interacting disordered systems (such as the transition between 
plateaux in the integer Quantum Hall or spin Quantum Hall effects). 
Although some similarity with what happens in systems with ordinary 
group symmetries is expected, it has also become clear over the last 
few years that more complicated scenarios are at work. Among the features
that play a key role in these scenarios are the facts that supergroup symmetries 
are not protected by the  
Mermin-Wagner theorem and can be spontaneously broken in two 
dimensions; that the lack of unitarity allows  non 
Kac-Moody  (logarithmic) conformal field theories with supergroup 
symmetry; etc. Recent references in this area are 
\cite{Zirnbaueri,Gruzberg,ReadSaleur,Victor,EFS}.

A systematic program to investigate these questions would be to tackle 
the continuum limit of spin chains with supergroup 
symmetries, either  with compact or non-compact representations,  
with or without integrability.
This turns out to be quite hard (for some recent work in this direction, 
see Refs.~\cite{Gade99,EFS}), as profoundly new features appear, on 
top of, and maybe related to, technical difficulties, already in the 
first steps of this program. 

\begin{figure}
\begin{center}
 \leavevmode
 \epsfysize=40mm{\epsffile{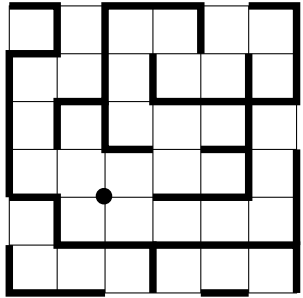}}
 \end{center}
 \protect\caption[3]{The configurations of the arboreal gas model are forests
of trees on the given graph; the union of trees jointly spans the lattice
and each component tree carries a Boltzmann weight $a$. As an example, we
show a forest on a portion of the square lattice. There are
six component trees, including an isolated vertex.}
\label{figforest}
\end{figure}

We discuss  in this paper one of the simplest possible examples of a 
supergroup theory one 
may imagine, the supersphere sigma model $OSP(1/2)/OSP(0/2)$. 
This model is deeply related to 
the arboreal gas model (see Fig.~\ref{figforest}), and the 
$Q\rightarrow 0$ limit of the Potts model on the square lattice, and 
through these relations, can be studied in considerable detail.  
In particular, the relation with the Potts 
model provides a phase diagram for the sigma model, a
confirmation that  the perturbative results are identical with those 
of the  $O(n),n=-1$ sigma model, 
and   the identification of `a' 
critical point of the sigma model
with the 
antiferromagnetic transition 
in the Potts model, as well as the transition in the arboreal gas 
model. Surprisingly, the associated 
critical theory turns out to be of a type seldom if ever encountered 
before in lattice models: with $c=-1$,  it involves a non-compact boson 
as well as a symplectic fermion, and is related to a theory with a larger, $OSP(2/2)$, 
symmetry. 

The paper is organized as follows. In section 2, we review the 
relationship between the $Q=0$ Potts model, the arboreal gas model, 
and the sigma model. A short version of the results presented there 
has appeared in Ref.~\cite{forests}. Related results have also appeared 
in Ref.~\cite{JSS}. Sections 3 and 4 discuss the continuum 
limit of the model at the antiferromagnetic critical point using Bethe 
ansatz equations as well as techniques from conformal field theory. In 
section 5 we prove the remarkable identity between the Bethe equations 
for the  $Q=0$ antiferro 
Potts model and the integrable $osp(2/2)$ spin chain in the 
fundamental representation. Using general results about the phase 
diagrams of orthosymplectic chains provides independent confirmation 
of our continuum limit identification. Some conclusions are gathered 
in section 6 while the appendix discusses a few technical aspects of 
Bethe ansatz for the $sl(2/1)=osp(2/2)$ algebra.

This paper is not only  a step up the program of understanding statistical 
mechanics models with supergroup symmetries: first, the  results will prove 
extremely useful in the analysis of the antiferromagnetic Potts model 
for $Q$ generic, which we will discuss elsewhere \cite{bigantiferro}; 
second, the arboreal gas is a model of great interest from the point 
of view of transitions in gases of `hard' objects at negative 
fugacity; finally, the relationship between the arboreal gas and 
the sigma model provides a way to investigate fundamental questions---like
the role of global topological properties in sigma models in 
their broken symmetry phase---in a particularly simple context.

\section{The $Q=0$ Potts model and the sigma model}

The Potts model at $Q=0$ can only be well defined through the cluster 
expansion and a proper limiting procedure \cite{SaleurAF}. This  requires 
generally to approach coupling $K=0$ along lines ${\rm e}^{K}-1={Q\over 
a}$, in which case the partition function as $Q\rightarrow 0$ can be 
rewritten as the partition function of an arboreal gas 
\cite{forests,JSS}
\begin{equation}
    Z_{\rm arboreal}=\sum_{p=1}^{\infty}a^p N(F_p)\label{arboreal},
\end{equation}
where $N(F_{p})$ is the number of ways of covering the lattice with 
$p$ non-overlapping trees (a forest), such that each site belong to one and 
only one tree. 

The proof of this statement is straightforward. Write the partition 
function as 
\begin{eqnarray}
    Z_{\rm Potts}&=&\sum_{\sigma_{i}=1,\ldots,Q}\prod_{<ij>}{\rm e}^{K\delta_{\sigma_{i}\sigma_{j}}}
    \nonumber\\
    &=&\sum_{G}\left({\rm e}^{K}-1\right)^{B}Q^{Cl}\nonumber\\
    &\approx&\left({Q\over a}\right)^{S}\sum_{G}a^{Cl-Cy}Q^{Cy}\label{zderiv}
\end{eqnarray}
where  $G$ are subgraphs of the lattice made of $B$ bonds and $Cl$ 
clusters (connected components), $S$ is the total number of sites, 
$Cy$ the number of cycles on the clusters. We have  used Euler's relation (valid in the plane or on an annulus, 
a geometry we will consider below), $S=B+Cl-Cy$. In the last equation $Q\approx 
0$, ${\rm e}^{K}-1\approx {Q\over a}$. As $Q\rightarrow 0$, only graphs 
without cycles survive, giving the result (\ref{arboreal}) after a 
renormalization of $Z$.  This is true irrespective of the lattice.

On the square lattice, the usual critical Potts model corresponds to 
approaching $Q=0$ along the selfdual curve ${\rm e}^K-1=\sqrt{Q}$
\cite{Baxsq}, i.e., by taking $a\rightarrow 0$, in which 
case one keeps a single spanning tree. The geometrical properties are 
then described in terms of a symplectic fermion theory 
\cite{DuplSal,DavDupl,Sal}. 

Similarly, on the triangular lattice the selfdual curve reads
$({\rm e}^K-1)^3+3({\rm e}^K-1)^2=Q$ \cite{Baxtri},
which, for $Q \to 0$ and $K>0$, becomes ${\rm e}^K-1 \approx \sqrt{Q/3}$,
whence $a\rightarrow 0$ and the equivalence with a spanning tree follows
once again. These two lattices thus strongly suggest that the usual
critical behaviour is universal, and governed by symplectic fermions.

The case $a>0$ corresponds to a massive region in the Potts model 
phase diagram, and therefore the arboreal gas is massive. The case 
$a<0$ negative is more interesting. While the model is still massive 
for $|a|$ big enough, on the square lattice and for 
$a^*<a\leq 0$, the model is massless, and flows 
to the symplectic fermion theory again. The special value $a^*=-4$ is 
critical, and corresponds to the limit $Q\rightarrow 0$ of the 
critical antiferromagnetic Potts model on the square lattice 
\cite{Baxter,SaleurAF}. 

On the triangular lattice there is strong numerical evidence
for a massless regime when $a^* < a \leq 0$,
again characterized by a flow to the symplectic fermion theory \cite{JSS}.
However, since $a^* = -5.705 \pm 0.006$ is only known numerically, it appears
to be difficult to detect whether there is distinct critical behaviour
exactly at $a^*$, and what is the exact nature of the transition to the
massive regime $a < a^*$ \cite{JSS}.

It turns out that, on any planar graph, the arboreal gas can be expressed in terms of an 
interacting symplectic fermion theory \cite{forests}. This comes from the remarkable 
identity
\begin{equation}
 \int {\rm d}\eta_1 {\rm d}\eta_2 \,
 {\rm e}^{a(Q_{1}-Q_{2})} {\rm e}^{\eta_1 M \eta_2} =
 \sum_{p=1}^{\infty}a^p N(F_p)\label{identity}
 \end{equation}
 Here,  each lattice site $i$  carries a pair of fermions $\eta_{1}(i)$ 
 and $\eta_{2}(i)$, $d\eta_{1}d\eta_{2}$ is shorthand for the measure 
 of integration over every site, and $M$ is 
 the discrete Laplacian with matrix elements $m_{ij}$. Thus, for $i\neq 
 j$, $m_{ij}$ equals minus the number of edges connecting $i$ and $j$,
 and $m_{ii}$ equals the number of neighbours of site $i$ (excluding $i$
 itself, if the graph contains a loop).
 We have   introduced the two objects
 \begin{eqnarray}
  Q_{1} &=& \sum_{i=1}^{S}\eta_{1}(i)\eta_{2}(i), \\
  Q_{2} &=& \sum_{\langle ij \rangle}
  \eta_{1}(i)\eta_{2}(i)\eta_{1}(j)\eta_{2}(j).
 \end{eqnarray}
where we recall $S$ is the total number of sites on the lattice, and the 
sum in $Q_{2}$ is taken over pairs of neighbouring sites.  This fermionic reformulation is interesting for several reasons, in 
 particular because it allows one to derive quick conclusions on the 
 phase diagram of the arboreal gas. One observes first that the 
 action in Eq.~(\ref{identity}) exhibits a non-linearly
 realized $OSP(1/2)$ symmetry. Indeed, let us introduce 
 on each vertex 
 an auxiliary field $\phi(i)$ subject to the constraint that 
 $\phi^{2}(i)+2a\eta_{1}(i)\eta_{2}(i)=1$. Solving for $\phi$ gives 
 two solutions, of which we chose $\phi=1-a\eta_{1}\eta_{2}$, that is 
 the `body' (part made of c-numbers) of $\phi$ positive (which we denote by 
 $b(\phi)>0$). Using the basic rule that
 $\delta({\phi^{2}+2a\eta_{1}\eta_{2}-1})={1\over 
 2\phi}\delta(\phi+a\eta_{1}\eta_{2}-1)$, together with ${1\over 
 \phi}={\rm e}^{a\eta_{1}\eta_{2}}$, one can entirely eliminate the $aQ_{1}$ 
 term in the action and write the functional integral as 
 \begin{eqnarray}
     {Z_{\rm arboreal}\over 2^{N}}&=&\int_{b(\phi)>0}
     \prod_{i}{\rm d}\phi(i){\rm d}\eta_{1}(i){\rm d}\eta_{2}(i) \,
     \delta[\phi^{2}(i)+
     2a\eta_{1}(i)\eta_{2}(i)-1]\nonumber\\
     & & \times \exp\left[\sum_{i,j}\eta_{1}(i)m_{ij}\eta_{2}(j) 
     -a\sum_{\langle ij \rangle}\eta_{1}(i)\eta_{2}(i)\eta_{1}(j)\eta_{2}(j)\right]\label{arbhemi}
 \end{eqnarray}
 We now introduce the vector $\vec{u}=(\phi,\eta_{1},\eta_{2})$ with the 
 scalar product 
 $\vec{u}(i)\cdot\vec{u}(j)=\phi(i)\phi(j)+a\eta_{1}(i)\eta_{2}(j)+a\eta_{1}(j)\eta_{2}(i)$.
 By explicitly writing down the discrete Laplacian terms and 
 regrouping a bit, one finds that the action can be rewritten as
 \begin{equation}
     S={1\over 
     a}\sum_{\langle ij \rangle}(\vec{u}(i)\cdot\vec{u}(j)-1)
 \end{equation}
 The additional 
 four-fermions term have disappeared from the action, which exhibits global 
 $OSP(1/2)$ symmetry (the group of transformations leaving the scalar 
 product invariant). The generating function in fact looks exactly 
 like the one of a discretized sigma model on the target manifold 
 $OSP(1/2)/OSP(0/2)\equiv UOSP(1/2)/SU(2)$! 

Although the  parameter $a$ appears in our definition of the scalar 
product, it 
 can always be absorbed in a rescaling of the fermions, and thus
 \begin{equation}
     {a^{N}Z_{\rm arboreal}\over 2^{N}}=\int_{b(\phi)>0}
     [{\rm d}\phi {\rm d}\eta_{1}{\rm d}\eta_{2}] \,
     \delta(\phi^{2}+2\eta_{1}\eta_{2}-1)\exp\left[-{1\over 
     a}\sum_{\langle ij \rangle}(\vec{u}(i)\cdot\vec{u}(j)-1)\right]
 \end{equation}
 where now 
 $\vec{u}(i)\cdot\vec{u}(j)=\phi(i)\phi(j)+\eta_{1}(i)\eta_{2}(j)+\eta_{1}(j)\eta_{2}(i)$. An infinitesimal
 $OSP(1/2)$ transformation reads
 \begin{eqnarray}
 \delta \phi&=&-\delta\xi_{1}\eta_{1}+\delta\xi_{2}\eta_{2}\nonumber\\
 \delta \eta_{1}&=&-\delta\xi_{2}\phi+\delta a\eta_{1}+\delta 
 c\eta_{2}\nonumber\\
 \delta \eta_{2}&=&-\delta\xi_{1}\phi+\delta b\eta_{1}-\delta a\eta_{2}
 \end{eqnarray}
 where $\delta\xi_{1},\delta\xi_{2}$ are `small' fermionic deformation 
 parameters, and $\delta a,\delta b,\delta c$ are small
 bosonic parameters.  In terms 
  of the fermion variables, the symmetry is thus  realized {\sl non-linearly}:
 \begin{eqnarray}
 \delta\eta_{1}=-\delta\xi_{2}(1-\eta_{1}\eta_{2})+\delta 
 a\eta_{1}+\delta c\eta_{2}\nonumber\\
 \delta\eta_{2}=-\delta\xi_{1}(1-\eta_{1}\eta_{2})+\delta 
 b\eta_{1}-\delta a\eta_{2}
 \end{eqnarray}
We now  take the 
 continuum limit by  introducing slowly varying  fermionic fields 
 $\eta_1(x,y)$ and $\eta_2(x,y)$. The euclidean action
 reads after some simple manipulations (Boltzmann weight ${\rm e}^{-S}$)
 \begin{equation}
 S={1\over g}\int 
    {\rm d}^{2}x \, \left[(\partial_{\mu}\phi)^{2}+
    2\partial_{\mu}\eta_{1}\partial_{\mu}\eta_{2}\right].
 \end{equation}
 with coupling $g=-2a$. Note that no lattice spacing (say, $a_{0}$) appears in this 
 action as the $a_{0}^{2}$ from the expansion of the fields is 
 compensated by the $1/a_{0}^{2}$ from the transformation of the 
 discrete sum over lattice sites into an integral. 

\begin{figure}
 \centering
     \includegraphics[width=2in]{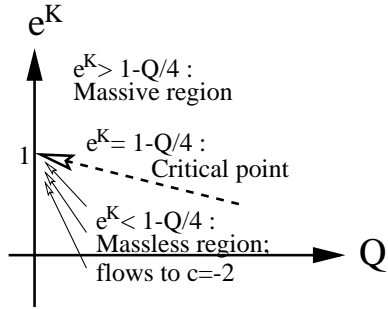}
 \caption{Phase diagram of the supersphere sigma model from the $Q$ 
 state Potts model on the square lattice. }
 \label{sigph}
 \end{figure}

We now go back to the issue of the sign of the  field $\phi$. 
There are of course {\sl two} solutions 
 to the equation $\phi^{2}+2\eta_{1}\eta_{2}=1$. While we have focused 
 on $\phi=1-\eta_{1}\eta_{2}$, the other choice $\phi=-1+\eta_{1}\eta_{2}$ 
 is also possible. It is therefore more correct to say we have mapped 
 the $Q\rightarrow 0$ Potts model or 
 the arboreal gas model onto a hemi-supersphere sigma model: the 
 correspondence with the sigma model 
 holds exactly at the perturbative level, but breaks 
 non-perturbatively.
 This is not expected to  affect the physics 
 in the broken symmetry phase where the field is slowly varying, and 
 does not jump between the two solutions. The physics in this phase 
 follows from the RG equation at 
 one loop:
 \begin{equation}
     {{\rm d}g\over {\rm d}l}=\beta(g)=(n-2)g^{2}
     \end{equation}
 with $n=-1=1-2$ for $OSP(1/2)$ and $g$  
 positive  (i.e., $a$ negative). We see that $g$ flows to $0$  in the IR for small 
 enough initial values, the theory remains massless, and the symmetry is 
 spontaneously broken.\footnote{When $g$ is negative (i.e., $a$ 
 positive), $g$ flows to large values in the IR, where the model is 
 massive, and the symmetry restored by the fluctuations.} This is  in
 complete agreement with the phase 
 diagram of the Potts model in the limit $Q\approx 0$, 
 ${\rm e}^{K}-1\approx {Q\over a}$ (see Fig.~\ref{sigph}). For the sigma model, the flow pattern
 holds until $g$ hits a critical value, 
 beyond which the model is massive in the IR again. On the square 
 lattice on the other hand, the massless phase of the Potts model 
ends up with  the 
 antiferromagnetic transition point, and  thus it is tempting to speculate 
 that the universality class of the critical sigma model coincides 
 with the critical antiferromagnetic Potts model. {\sl But} this might 
 be affected by the difference between  the hemi-supersphere and the full 
 one (the effect of global topological properties of manifolds on 
 the critical points of sigma models does not seem to be completely 
 understood in general. See \cite{Mouhanna} for a recent review.)
 
While the situation is not so familiar in two dimensions where most 
effort has been devoted to models based on unitary compact groups 
where the Mermin-Wagner-Coleman theorem prevents spontaneous symmetry 
breaking, it is well known in higher dimensions. There, one typically 
tackles exponents at the critical point in the $O(N)$ model with
$(\vec{\Phi}^{2})^{2}$ coupling by studying the broken symmetry phase, i.e.,
the non-linear sigma model, in $d=2+\epsilon$, calculating the $\beta$ 
function in this phase, and identifying the critical point with the 
first zero of the $\beta$ function. The perturbative series thus 
obtained are only sensitive to the local structure of the group 
manifold, while the restoration of symmetry involves global 
properties.

To discuss things further, we consider the geometrical interpretation
of the full sigma model. It is slightly more convenient to start from
the non-scaled fermions, and consider the equivalent of
Eq.~(\ref{arbhemi}) with the constraint $b(\phi)>0$ removed.  To each
site $i$ we assign an Ising variable $\epsilon_i = \pm 1$ which
distinguishes between the two solutions (i.e., $b(\phi)>0$ and $b(\phi)<0$).
One finds easily
  \begin{eqnarray}
     Z_{\rm sigma} &\equiv& \int
      \prod_{i}{\rm d}\phi(i){\rm d}\eta_{1}(i){\rm d}\eta_{2}(i) \, \delta[\phi^{2}(i)+
      2a\eta_{1}(i)\eta_{2}(i)-1]
      \, \exp\left[ -\frac{1}{a} \sum_{\langle ij \rangle}
      \left( \vec{u}(i) \cdot \vec{u}(j) - 1 \right) \right] \nonumber\\
     &=& \int \prod_{i}{\rm d}\eta_{1}(i){\rm d}\eta_{2}(i) \, 
      \sum_{\{\epsilon_{i}\}}\exp\left[-{1\over 
      a}\sum_{<ij>}(\epsilon_{i}\epsilon_{j}-1)\right] \\
      & & \times\exp\left[\sum_{i,j}\eta_{1}(i)m_{ij}\eta_{2}(j)+\sum_{i}[\eta_{1}(i)\eta_{2}(i)(a+\epsilon_{i}\tilde{\epsilon}_{i}-m_{ii})
      -a\sum_{\langle ij \rangle}\epsilon_{i}\epsilon_{j}\eta_{1}(i)\eta_{2}(i)\eta_{1}(j)\eta_{2}(j)\right] \nonumber
      \end{eqnarray}
where we have defined $\tilde{\epsilon}_{i}=-\sum_{j\neq i} m_{ij}
\epsilon_{j}$.  For the region of broken symmetry, $a$ is negative,
and thus the existence of domain walls with
$\epsilon_{i}\epsilon_{j}=-1$ extremely unfavorable for small $|a|$.
If we think of the $\epsilon_{i}$'s as Ising variables (essentially
the c-number parts of the field $\phi$), we have a ferromagnetic Ising
model coupled to the arboreal gas. A full geometrical interpretation
is easily obtained by going over the steps of \cite{forests}. The
$\eta$ terms again give an arboreal gas, but now each tree gets a more
complicated weight
 \begin{equation}
     a\rightarrow a + \sum_{i {\rm \ in \ tree}}
     (\epsilon_{i}\tilde{\epsilon}_{i}-m_{ii})-a\sum_{\langle ij \rangle
     {\rm \ in \ tree}} (\epsilon_{i}\epsilon_{j}-1)
 \end{equation}
 so now trees getting near or across domain walls have their weight 
 modified. 
 
 As the lattice coupling is modified, a first sight analysis would say 
 that generically in such a model, one will observe two transitions:
 an Ising transition where only the dynamics of the $\epsilon_{i}$ variables 
 becomes critical, and an arboreal gas
 transition, where only the dynamics of the trees becomes critical. 
These transitions  will occur for different values of $a$. Hence the 
 critical point of the arboreal gas, i.e., the hemi-sphere lattice sigma 
 model, will coincide in general with the critical point 
 of the coupled system [Ising model + arboreal gas], i.e., the 
 full lattice sigma model.\footnote{This is if the Ising transition has 
 not yet occurred when the arboreal degrees of freedom become critical. 
 The converse might  happen, with the Ising variables becoming 
 critical first, but this  seems quite unlikely, as it would 
  lead to the universality class of the supersphere sigma 
 model being a simple Ising theory. Notice that at small 
  negative $a$ the Ising model is in the ordered phase, and the weight 
  per unit length of a domain wall is ${\rm e}^{2/a} \ll 1$. As $|a|$ 
  increases, this weight increases until at the critical value $a=-4$ 
  it reaches ${\rm e}^{-1/2}=0.606$. Without coupling to the arboreal gas, 
  the Ising model meanwhile would order when the weight per unit length of a 
  domain wall is ${1\over 1+\sqrt{2}}=0.414$. So the Ising transition 
  must  be significantly hampered by the trees.} Only an extraordinary coincidence 
 could make the Ising and arboreal transitions merge, resulting in a higher order 
 phase transition, and a situation where the critical points of the 
 hemi-sphere and full sigma model are different, the latter being of 
 higher order, and less stable. 
 
 This  whole analysis may have to be taken with a grain of salt however. In 
 particular, the arboreal gas is a model whose weights are extremely 
 non-local, and the modification brought by coupling to the Ising 
 variables might have strange consequences. Also, it is after all not 
 clear how much of the conventional wisdom derived from models with 
 positive Boltzmann weights will  extend to models such as ours, where 
 the physics is determined by balance of positive and negative 
 contributions in addition to the usual energy/entropy balance. At 
 this stage,
 numerical study seems the best way to clarify the situation.

\section{Bethe equations at the antiferromagnetic point}

The special value $a=-4$ identified above corresponds more precisely  to 
the end point $Q\rightarrow 0$ of the antiferromagnetic 
critical line, which was identified by Baxter \cite{Baxter} as 
\begin{equation}
    \left({\rm e}^{K_{1}}+1\right)\left({\rm e}^{K_{2}}+1\right)=4-Q\label{antiferro}
\end{equation}
(here $K_{1},K_{2}$ are the usual horizontal and vertical couplings). 
While the Potts model on the square lattice is not solvable for 
arbitrary couplings, it does become solvable on the line 
(\ref{antiferro}). The Bethe ansatz equations were written in the original paper by 
Baxter, and read
\begin{equation}
    \left({\sinh (\lambda-i\gamma)\over 
    \sinh(\lambda+i\gamma)}\right)^{N}=\prod_{\lambda'}{\sinh{1\over 
    2}(\lambda-\lambda'-2i\gamma)\over \sinh{1\over 
    2}(\lambda-\lambda'+2i\gamma)}
\end{equation}
where $N$ is the width of the cylinder in the transfer matrix
(measured in terms of the Potts spins), the 
$\lambda$ are the usual spectral parameters determining the Bethe 
wave function, and we have parameterized $\sqrt{Q}=2\cos\gamma$, $\gamma\in\left[0,{\pi\over 2}\right]$. 
As discussed in details in \cite{bigantiferro} for the case $Q$ generic, the 
physics of the isotropic model $K_{1}=K_{2}$ can be studied more 
conveniently by considering a strongly anisotropic limit and the 
associated quantum hamiltonian for a chain of $2N$ spins $1/2$. The 
energy (in the relativistic scale) is then given by
\begin{equation}
    E={\pi-2\gamma\over\pi}\sum_{\lambda}{2\sin 
    2\gamma\over\cosh2\lambda-\cos 2\gamma}
\end{equation}
We expect that the ground state is obtained by filling 
up the lines Im $\lambda={\pi\over 2}$ and Im $\lambda=-{\pi\over 2}$. 
If we restrict to solutions of the type $\lambda\equiv {i\pi\over 
2}+\alpha$ and $\lambda\equiv -{i\pi\over 
2}+\beta$, a system of two equations for the real parts $\alpha$ 
and $\beta$ can then we written. It can also be extended to string 
configurations centered on either line. 

We now focus on the limit $\gamma\rightarrow {\pi\over 2}$. We set 
$\gamma={\pi\over 2}+\epsilon$, and scale the spectral parameters 
with $\epsilon$. The resulting system of equations reads (we use the 
same notations for the rescaled spectral parameters)
\begin{eqnarray}
    \left({\alpha-i\over\alpha+i}\right)^{N}=\prod 
    {\alpha-\beta-2i\over\alpha+\beta+2i}\nonumber\\
    \left({\beta-i\over\beta+i}\right)^{N}=\prod 
	 {\beta-\alpha-2i\over\beta-\alpha+2i}\label{bethe}
\end{eqnarray}
and the energy 
\begin{equation}
    E=-{4\over\pi}\sum {1\over 1+\alpha^{2}}+{1\over 1+\beta^{2}}
\end{equation}

\section{Critical properties at the antiferromagnetic point}

\subsection{First aspects}

We now observe that setting $\alpha=\beta$ 
gives Bethe  equations similar to the 
 antiferromagnetic XXX chain, while there is an 
additional factor of $2$ in the definition of the energy. The sector 
symmetric in $\alpha$ and $\beta$ roots will 
 lead to {\sl twice} the central charge and critical exponents of the 
 XXX chain, which are the ones of a free boson with coupling constant 
 $g=1$, or an $SU(2)$ WZW model at level one (a similar feature was 
 encountered in \cite{EFS}. This is more than a coincidence, as will 
 be discussed below).

This is not to say that all the excitations 
are of this type. Doubling the spectrum of the XXZ chain of course 
creates many gaps in the conformal towers, which have to be filled 
with other types of excitations in our system. But at least, we can 
derive quickly some results from this subset of the spectrum, which 
is now under control. 

For the electric part of the spectrum, we then get the following 
result
\begin{equation}
    c=2\times \left(1-{6e^{2}\over g}\right)=2-12 e^{2}
\end{equation}
The charge $e$ can be 
interpreted as giving a special weight  $w$ to  non-contractible loops. 
Following the detailed mappings of the Potts model on contours and 
vertex models, one finds  $w=2\cos\pi e$. 
Since non-contractible loops come in pairs, we have $e$ defined 
modulo an integer. The $Q\rightarrow  0$ state Potts model itself corresponds to 
$e=e_{0}={1\over 2}$, for which $c=-1$, while the effective central 
charge is $c_{\rm eff}=2$. 

Magnetic gaps read similarly
\begin{equation}
    \Delta=2 \times {gm^{2}\over 4}={m^{2}\over 2}
\end{equation}
Here $m$ is the spin of the equivalent XXX chain. Since one 
solution of the equivalent XXX system corresponds to a {\sl pair} of 
solutions of our system,  $m={N\over 2}-[\hbox{number of pairs of 
solutions}]$ and $m={S^{z}\over 2}$, where $S^{z}$ is the spin in our initial 
system. Therefore 
\begin{equation}
    \Delta={(S^{z})^2\over 8}
\end{equation}
where  $S^{z}$ is an integer since 
we have an even number of 
spins.

One can see here that parity effects will likely occur.
 If $N$ is even, since we must also have  an even number of 
 upturned spins to make up pairs of Bethe roots in the ``folding'' of 
 the equations, the result 
 strictly speaking will only apply to $S^{z}$ even,  including the 
 ground state. Numerical study shows that in fact it holds for 
 $S^{z}$ odd as well. If $N$ is odd on the other hand, 
 since we must also have  an even number of 
 upturned spins to make up pairs of Bethe roots in the ``folding'' of 
 the equations, the result 
 strictly speaking will only apply to $S^{z}$ odd. Numerically, we 
 have observed indeed that for $S^{z}$ even, including the ground 
 state $S^{z}=0$, results for $N$ odd obey a different pattern.  This 
 will be discussed later, and we restrict for now to $N$ even.
 
 The electric charges in the $S^{z}=0$ sector of the equivalent XXX 
  chain are, for the untwisted 
  case,  integer, and give rise to the following gaps $\Delta=2\times 
  {e^{2}\over 4g}={e^{2}\over 2}$. Combining all the excitations, we 
  thus get  
  exponents from a {\sl single} free boson at $g={1\over 2}$, viz.,
  $\Delta_{em}={1\over 2}\left(e\pm {S^{z}\over 2}\right)^{2}$,
  where $e,S^{z}$ are integers. 
  \footnote{In general, we will denote by $\Delta$ the (rescaled) gaps 
  with respect to the central charge $c=2$, and reserve the notation 
  $h$ for the (rescaled) gaps with respect to the true ground state of 
  the antiferromagnetic Potts model.} This is intuitively 
  satisfactory, as one expects the ``arrows degrees of freedom'' in the 
  model to be described by a free boson in general.  This free boson of course 
  contributes only $c=1$ to the total central charge, so to get $c=2$ we 
  need to invoke the presence of another theory with $c=1$. 
  
  To proceed, we consider the dynamics of the hole excitations over 
  the ground state where the two lines $\hbox{Im }\lambda=\pm 
  {\pi\over 2}$ are filled. 
  We introduce the functions (notations are the same as in 
  \cite{Saleurslnk})
\begin{equation}
   e_{n}(\alpha)={\alpha+in/2\over 
   \alpha-in/2};~~~a_{n}(\alpha)={i\over 2\pi} {{\rm d}\over {\rm d}\alpha} 
   \ln\left[e_{n}(\alpha)\right]
\end{equation}
and the Fourier transform
\begin{equation}
\hat{f}(x)=\int {\rm d}\alpha \, {\rm e}^{i\alpha x}f(\alpha)
\end{equation}
such that $\hat{a}_{n}(x)={\rm e}^{-n|x|/2}$. We take the logarithm of the Bethe 
equations [Eq.~(\ref{bethe})] and introduce densities of particles ($\rho$) and holes 
($\rho^{h}$) per unit length. The result in  Fourier transforms reads
\begin{eqnarray}
   \hat{\rho}_{1}+\hat{\rho}_{1}^{h}=a_{2}-a_{4}\star\rho_{2}\nonumber\\
   \hat{\rho}_{2}+\hat{\rho}_{2}^{h}=a_{2}-a_{4}\star\rho_{1}	
\end{eqnarray}
and is of the form $\hat{\rho}+\hat{\rho}^{h}=\hat{p}+K\hat{\rho}$. Physical 
equations  (that is, equations where the right hand side depends on 
density of excitations) meanwhile read
\begin{eqnarray}
   \hat{\rho}_{1}+\hat{\rho}_{1}^{h}={1\over 2\cosh x}
   +{1\over 1-{\rm e}^{4|x|}} \hat{\rho}_{1}^{h}+
   {1\over 2\sinh 2x}\hat{\rho}_{2}^{h}\nonumber\\
   \hat{\rho}_{2}+\hat{\rho}_{2}^{h}={1\over 2\cosh x}
   +{1\over 2\sinh 2x}\hat{\rho}_{1}^{h}
  +{1\over 1-{\rm e}^{4|x|}} \hat{\rho}_{2}^{h}
\end{eqnarray}
Introducing the matrix $Z=(1-K)^{-1}$ one finds
\begin{equation}
   Z=\left(\begin{array}{cc}
   {1\over 1-{\rm e}^{-4|x|}} &{-1\over 2\sinh 2x}\\
   {-1\over 2\sinh 2x}&{1\over 1-{\rm e}^{-4|x|}}\end{array}\right)_{x=0}
   \end{equation}
Now in ordinary models such as those with $sl(n)$ symmetry, it is well 
known how to obtain analytically the spectrum of exponents by careful 
analysis of finite size scaling effects.  The exponents are expressed 
as quadratic forms based on the matrix $Z$ and its inverse 
\cite{SuzJ,DeVega}. We shall 
assume a bit naively that the same essentially is true here as well. 
All the elements of $Z$ diverge in the limit $x=0$, but if we keep $x$ small,
$Z$ reads
\begin{equation}
   Z=\left(\begin{array}{cc}
    {1\over 2}+{1\over 4\epsilon} & {-1\over 4\epsilon}\\
   {-1\over 4\epsilon} & {1\over 2}+{1\over 4\epsilon}\end{array}\right)
   \end{equation}
which allows us to calculate the inverse matrix as 
$\epsilon\rightarrow 0$
\begin{equation}
    Z^{-1}=\left(\begin{array}{cc}
    1&1\\
    1&1\end{array}\right)
    \end{equation}
We can now let $\epsilon\rightarrow 0$ to express the finite size 
scaling corrections and thus the gaps  for $N$ even as \cite{DeVega}
\begin{equation}
    x={1\over 4} \delta n^{t}Z^{-1}\delta n+d^{t}Zd
\end{equation}
where $\delta n$ is the two-column vector $\left(\begin{array}{c}
\delta n_{1}\\
\delta n_{2}
\end{array}\right)$, $\delta n_{i}$ the change of solutions of the 
$i^{th}$ type. Similarly, 
$d$ is the two-column vector $\left(\begin{array}{c}
d_{1}\\
d_{2}
\end{array}\right)$, $d$ the number of particles backscattered from 
the left to the right of the 
Fermi sea. Therefore we find
\begin{equation}
   \Delta+\bar{\Delta}= {1\over 4}(\delta n_{1}+\delta n_{2})^{2}+0\times 
   (\delta n_{1}-\delta 
   n_{2})^{2}+{1\over 4}(d_{1}+d_{2})^{2}+\infty \times 
   (d_{1}-d_{2})^{2}\label{spectrum}
\end{equation}
The symmetric part of the spectrum $\delta n_{1}=\delta n_{2}=\delta 
n$ and $d_{1}=d_{2}=d$ gives $h+\bar{h}=(\delta n)^{2}+d^{2}$. We have 
checked numerically that this formula extends to the case $\delta n$ 
half odd integer, so the total XXX part of the spectrum reads
$h+\bar{h}=e^{2}+{m^{2}\over 4}$, $e,m$ integers, or $h={1\over 
8}(m+2e)^{2}$ and $\bar{h}={1\over 8}(m-2e)^{2}$. This 
corresponds to a free boson at the coupling $g=1/2$, i.e., symplectic 
fermions. It is useful to introduce now the determinants of the 
Laplacian on a torus with different types of boundary conditions, 
$\hbox{Det }\Delta_{ab}$;
the subindices denote the boundary conditions
($P =$ periodic, $A =$ antiperiodic, or $F =$ free), with $b$ referring to
the space direction (of length $N$) 
and $a$ to the imaginary time direction (of length $M$, with 
modular parameter $q={\rm e}^{-2\pi M/N}$) \cite{yellow}:
\begin{eqnarray}
   \hbox{Det }\Delta_{AP}&=&{1\over \eta\bar{\eta}} 
   \sum_{n,m} q^{(2n+1)^{2}/8}\bar{q}^{(2m+1)^{2}/8}
   \nonumber\\
   \hbox{Det }\Delta_{AA}&=&{1\over \eta\bar{\eta}} 
       \sum_{n,m} q^{(2n)^{2}/8}\bar{q}^{(2m)^{2}/8}\nonumber\\
       \hbox{Det }\Delta_{PA}&=&{1\over \eta\bar{\eta}} 
	   \sum_{n,m} (-1)^{n}q^{(2n)^{2}/8}(-1)^{m}
	   \bar{q}^{(2m)^{2}/8}
\end{eqnarray}
where  $\eta=q^{1/24}\prod_{1}^{\infty}(1-q^{n})$ is Dedekind's eta 
function. The generating function of  XXX levels in the case $N$ even thus reproduces 
$\hbox{Det }\Delta_{AP}+\hbox{Det }\Delta_{PA}
+\hbox{Det }\Delta_{AA}$.\footnote{Notice that since $\hbox{Det }\Delta_{PP}=0$, 
we can write as well 
$\hbox{Det }\Delta_{AP}=
{2\over \eta\bar{\eta}} 
       \sum'_{n,m} q^{(2n+1)^{2}/8}\bar{q}^{(2m+1)^{2}/8}$,
where the sum is over $m$ and $n$ of the same parity.}

Combined with this symplectic fermion part, we have an infinite 
degeneracy associated with the modes $d_{1}=d_{2}$. Presumably, this 
leads to a continuous spectrum of exponents on top of the symplectic 
fermion ones. Since on the other hand we know that the theory has 
$c=-1$ ($c_{\rm eff}=2$) while the symplectic fermion part identified so 
far has $c=-2$ ($c_{\rm eff}=1$) , it is very tempting at this stage to 
propose that the missing part of the spectrum consists simply in a 
non-compact boson. We now give a more direct argument for this.

\subsection{Lessons from free boundary conditions and modular 
transformations}

The partition function of the $Q=0$ Potts model on an annulus with 
free boundary conditions can be obtained in two steps, 
both based on previous works. The first step consists is expressing 
it as an alternate sum over generating functions of levels for type II 
representations of $U_{t}sl(2)$ (where 
$\sqrt{Q}=t+t^{-1}$, and we use the notation $t$ for the quantum 
deformation parameter to avoid confusion with $q$) \cite{PasquierSaleur} in the associated six-vertex or XXZ 
model (not to be confused with the XXX model which appears through 
its Bethe equations earlier in this paper as a subsector of the Potts 
model. The latter has $t=1$ while the former has $t=i$). The second 
step consists in identifying the generating function $K_{j}$ of levels for 
the type II representations. This step is not rigorous, and based on 
our understanding of the model at special points $\gamma={\pi\over 
n}$, $n>2$, together with arguments of continuity in $\gamma$ 
\cite{SaleurAF}. To 
make a long story short, one ends up with the following results.

The expression of $K_{j}$ as $Q\rightarrow 0$ is 
\begin{equation}
   K_{j}={q^{j(j+1)/2+1/24}\over P^{2}(q)}
   \left[ 1+2\sum_{n=1}^{\infty}q^{2n^{2}+n(2j+1)}-2
   \sum_{n=0}^{\infty}q^{2(n+1/2)^{2}+(n+1/2)(2j+1)}\right]
\end{equation}
where $P(q)=\prod_{1}^{\infty}(1-q^{n})$, and $j$ is an integer (spin) label, $j=0,1,2\ldots$. 
The astute reader might wonder why this expression does not satisfy 
for instance $K_{0}-K_{1}+K_{2}-K_{3}\ldots=0$ which would naively result from 
the quantum group symmetry that mixes all spin $0$ and spin $1$ 
representations \cite{PasquierSaleur}. The reason is that the eigenvalues of the transfer 
matrix all go to zero as $Q\rightarrow 0$, and this transfer matrix cannot 
be rescaled in such a way that its elements   remain all finite as 
$Q\rightarrow 0$. Alternatively, in a hamiltonian formulation, the 
ground state energy of the relativistic hamiltonian goes to $-\infty$ 
in the limit $Q\rightarrow 0$. In that limit, the $K_{j}$'s are 
obtained after an additional renormalization, and do not have to 
satisfy the quantum group induced coincidences any longer.

A little algebra  shows that the generating function  of levels (partition 
function  in the conformal language) of the XXZ or six-vertex  
model on the annulus is 
\begin{equation}
   Z_{PF}^{\rm vertex}=\sum_{j=0}^{\infty}(2j+1)K_{j}={q^{1/24}
   \over P^{2}(q)}\sum_{N=0}^{\infty }
   q^{N(N+1)/2}
\end{equation}

In the usual case of the critical $Q=0$ state Potts model, the 
equivalent of $K_{j}$ would be 
\begin{equation}
    \tilde{K}_{j}={q^{j(j-1)/2+2/24}\over P(q)}\left(1-q^{1+2j}\right)
\end{equation}
(in that case one checks that 
$\tilde{K}_{0}-\tilde{K}_{1}+\tilde{K}_{2}-\tilde{K}_{3}\ldots=0$ indeed).
The partition function of the vertex model 
would then be in that case
\begin{equation}
    \tilde{Z}^{\rm vertex}_{PF}=4{q^{2/24}\over P(q)}\sum_{N=0}^{\infty} q^{N(N+1)/2}
\end{equation}
It corresponds to the partition function of symplectic fermions on an 
annulus with antiperiodic boundary conditions in the time direction,
$\hbox{det}\left(-\Delta_{AF}\right)$, and 
is as well a 
character of the $c=-2$ rational logarithmic theory 
\cite{FlohrBredthauer}. The object $Z^{\rm vertex}_{PF}$ written as 
$Z^{\rm vertex}_{PF}=\tilde{Z}^{\rm vertex}_{PF}\times {1\over \eta}=
\hbox{det}\left(-\Delta_{AF}\right)\times {1\over \eta}$ can thus be interpreted as 
arising from a theory made of symplectic fermions and a non-compact 
boson, with partition function ${1\over \eta}$. 

The presence of the non-compact boson can be ascertained through a 
modular transformation. Writing 
$\tilde{Z}^{\rm vertex}_{PF}=2{\theta_{2}(q)\over \eta(q)}$, we see that 
modular transformations in the usual $Q=0$ critical Potts model do  close, since 
\begin{eqnarray}
    \theta_{2}(-1/\tau)&=&\sqrt{-i\tau}\theta_{4}(-1/\tau)\nonumber\\
    \theta_{4}(-1/\tau)&=&\sqrt{-i\tau}\theta_{2}(-1/\tau)\nonumber\\
	\eta(-1/\tau)&=&\sqrt{-i\tau}\eta(\tau)
\end{eqnarray}
In the antiferromagnetic case meanwhile
we now generate a continuous 
spectrum, according to
\begin{equation}
    {1\over \eta(-1/\tau)}\propto \int {\rm d}\alpha \, 
    {{\rm e}^{2i\pi\tau\alpha^{2}}\over\eta(\tau)}
\end{equation}

As for the arboreal gas or (the properly defined limit of) the Potts 
model itself, the partition function is
\begin{equation}
    Z^{\rm arboreal}_{PF}=\sum_{0}^{\infty}(-1)^{j}K_{j}={q^{1/24}\over  P^{2}}\sum_{N=0}^{\infty} 
    (2N+1)(-1)^{N}q^{N(N+1)/2}
\end{equation}
This can be written, using the Jacobi identity, as  
\begin{equation}
    Z^{\rm arboreal}_{PF}=q^{1/24}P(q)
\end{equation}
The corresponding object 
in the usual critical  Potts model ($\tilde{Z}_{PF}$) would vanish identically. As 
before, $Z^{\rm arboreal}_{PF}$ 
can  be interpreted as the partition function of symplectic 
fermions (here, fermions periodic on the annulus, and the zero mode 
subtracted) times the contribution from a non-compact boson, ${1\over \eta}$. 

%
%
%
%
%

We thus see that the study of the model with free boundary conditions 
greatly strengthens the claim that the 
continuum limit of the model is made up of a symplectic 
fermion theory and a non-compact boson. This of course agrees with 
$c=-1$, $c_{\rm eff}=2$. Let us discuss this identification further.

\subsection{The continuum limit of the $Q=0$ antiferromagnetic 
critical Potts 
model}

The action corresponding to this continuum limit reads (Boltzmann weight
${\rm e}^{-S}$)
\begin{equation}
    S\propto \int {\rm d}^{2}x \, 
    \left[\left(\partial_{\mu}\phi\right)^{2}+2\partial_{\mu}\eta_{1}\partial_{\mu}\eta_{2}
    \right]\label{decaction}
\end{equation}
with the boson $\phi$ being non-compact. 
Now  recall that the scaling limit of the arboreal gas, 
or the  $Q=0$ Potts model in the Berker-Kadanoff phase, 
was identified with the $OSP(1/2)$ supersphere sigma model, whose 
action reads exactly like Eq.~(\ref{decaction}) with the 
additional constraint that $\phi^{2}+2\eta_{1}\eta_{2}=1$. To all 
appearances therefore, the critical point of the arboreal gas simply 
corresponds to a free $OSP(1/2)$ field theory, where the constraint 
has been forgotten, and  the symmetry is 
realized linearly. This interpretation requires some further discussion. 
The sigma model in the massless Berker-Kadanoff (spontaneously broken symmetry) phase 
has action 
\begin{equation}
    S={1\over g}\int {\rm d}^{2}x \, \left[\left(\partial_{\mu}\phi\right)^{2}+
    2\partial_{\mu}\eta_{1}\partial_{\mu}\eta_{2}
    \right]
\end{equation}
with $g$ a {\sl negative} quantity. It thus looks as if the natural 
action for the theory without constraint should be Eq.~(\ref{decaction}) 
with a negative sign. While this has no importance for the fermions 
(the sign for them can be switched by a trivial relabeling) it is 
not so at all for the boson. While the model  $S=+\int 
(\partial\phi)^{2}$  is 
well defined, the model with $S=-\int 
(\partial\phi)^{2}$ is naively rather ill defined! It is possible to 
argue however, that since the theory is not compactified, this sign 
does not really matter. 

Start with the `proper' free uncompactified boson, with action (we put 
numerical
constants explicitly now)
\begin{equation}
    S={1\over 2}\int {\rm d}^2 x \, \partial_{\mu}\phi\partial_{\mu}\phi
\end{equation}
and thus propagator $\langle \phi(z,\bar{z})\phi(0)\rangle =-{1\over 
4\pi}\ln |z|^{2}$. The fact that $\phi$ is not compactified allows all 
powers of $\phi$ as fields, and hence by summation, all complex 
exponentials. However, since only 
normalizable states appear in the partition function (see 
\cite{Gawe,Saleur} for further discussion on this point), only imaginary 
exponentials will show up there. This is compatible with the result of the 
functional integration
\begin{equation}
    Z={1\over\sqrt{Im\tau}} {1\over \eta\bar{\eta}}={1\over 
    \eta\bar{\eta}}\int_{-\infty}^{\infty} (q\bar{q})^{ 
    \alpha^{2}/4} \, {\rm d}\alpha\label{noncompactz}
\end{equation}
Hence in a transfer matrix study of such a model, we would see essentially a 
continuous spectrum above the central charge $c=1$. Note how the 
presence of operators with arbitrarily negative dimension does not 
make the theory unstable,  nor does it affect the ground state in the 
transfer matrix calculations.

If we admit that the ``full operator content'' is built out of fields 
${\rm e}^{(a+ib)\phi}$ and derivatives, then a change of variables $\phi\rightarrow 
i\theta$ 
should leave it invariant. This means that full operator content can 
be built as well out of ${\rm e}^{(a+ib)\theta}$ and derivatives, where now 
$\theta$ has a propagator with the wrong sign, $\langle 
\theta(z,\bar{z})\theta(0)\rangle =-{1\over 
4\pi}\ln |z|^{2}$. Normalizable states now correspond to real 
exponentials, and their contribution to the partition function gives 
exactly the same result as Eq.~(\ref{noncompactz}), which can therefore 
be considered as the partition function for 
the theory with  action
\begin{equation}
S=-{1\over 2}\int 
{\rm d}^2x \, \partial_{\mu}\theta\partial_{\mu}\theta\label{noncompactzz}
\end{equation}
In other words, we define the functional integral in the case 
of Eq.~(\ref{noncompactzz}) by analytic continuation, and since the set of 
allowed fields is invariant under this continuation, the partition function is the same in 
both cases. 

The fact that the critical point of the supersphere sigma model 
appears to be the free theory is a bit surprising. One would have 
expected instead the $OSP(1/2)$ symmetry to be restored in a non-trivial
way, and the model maybe to coincide with  the critical  
$O(n=-1)$ model of Nienhuis  
 \cite{Nienhuis}, discussed more thoroughly as an $OSP(1/2)$ model in Ref.~\cite{ReadSaleur}. The 
latter model however has central charge $c=-{3\over 5}$, and 
definitely is very different from the one we obtain.  This difference 
might be related to the difference between the hemi-supersphere and 
the full supersphere sigma model.  It might also 
be that  some of the simplifications in the definition of the  
solvable $O(n)$ 
critical model in \cite{Nienhuis} (such as the definition of the Boltzmann weight) affect 
the physics in an unexpected way when $n$ is negative, with, as a 
result, less universality than expected.

\section{Relation with $OSP(2/2)$ theories}

\subsection{$osp(2/2)$ integrable spin chain.}

We now notice that the Bethe  equations bear a strange similarity 
with  equations written for a 
chain with $sl(2/1)$ symmetry. In particular, if we took 
a  chain with alternating $3,\bar{3}$ representations ($N$ of either 
type), the equations 
would read, in the appropriate grading \cite{Links,EFS},
\begin{eqnarray}
	 \left({\alpha-2i\over\alpha+2i}\right)^{N}=\prod 
	 {\alpha-\beta-2i\over\alpha+\beta+2i}\nonumber\\
	 \left({\beta-2i\over\beta+2i}\right)^{N}=\prod 
	      {\beta-\alpha-2i\over\beta-\alpha+2i}
\label{bae33chain}
     \end{eqnarray}
More generally, the Bethe equations for alternating representations build 
with $n$-supersymmetric tensors and the conjugate would 
look like Eq.~(\ref{bae33chain}), but with the factor $2$ replaced by $2n$ on
the left-hand side (source terms).  Our equations seem to 
correspond to taking ``quarter spin'' 
representations (in $sl(2)$ inspired conventions, where the 
fundamental has spin $1/2$). This in fact has a meaning, thanks to 
the work \cite{PfannFrahm} where it was shown 
that the Bethe equations for a chain 
built with the four-dimensional typical representation (and $L$ sites) read
\begin{eqnarray}
	 \left({\alpha-i(2b+1)\over\alpha+i(2b+1)}\right)^{L}=\prod 
	 {\alpha-\beta-2i\over\alpha+\beta+2i}\nonumber\\
	 \left({\alpha+i(2b-1)\over\alpha-i(2b-1)}\right)^{L}=\prod 
	      {\beta-\alpha-2i\over\beta-\alpha+2i}\label{frahmbethe}
     \end{eqnarray}
where $b$ is the fermionic Dynkin number (denoted by $b_{2}$ in 
\cite{PfannFrahm}). The fundamental of $sl(2/1)$ corresponds 
to $b=-1$, a degenerate case. The case $b=0$ meanwhile 
corresponds to the fundamental 
representation of $osp(2/2)$. Energy terms in this case can also be 
read from \cite{PfannFrahm} and exactly match the term in the $Q=0$ 
Potts model case (after a rescaling of the Bethe parameters in 
\cite{PfannFrahm}). We thus conclude that the Bethe equations and 
energy for the $Q=0$ antiferromagnetic Potts spin chain are identical with 
those for the integrable spin chain with $osp(2/2)$ symmetry built 
on the four-dimensional fundamental representation $[b=0,j=1/2]$. 

Note that the length of the $osp(2/2)$ system in these equations, 
denoted $L$ in the Bethe equations, coincides with $N$, where the 
number of spin $1/2$ representations in the spin chain is $2N$. Hence 
the size of the Hilbert spaces on which the two systems are acting are 
identical, and equal to $2^{2N}=4^{N}$. Since the Bethe equations also 
coincide, this is a very strong indication that the spectra of the 
two systems are identical, even though one has to be careful with 
such statements, as Bethe spectra are not always complete in the 
superalgebra case.\footnote{Note that such  a `transmutation' from 
Bethe equations characteristic of a rank-one algebra to those of a 
rank-two algebra was met already 
in 
the $sl(3)$ regime of the $a_{2}^{(2)}$ chain \cite{Tsvelik}. }

\subsection{Continuum limit of the $osp(2/2)$ integrable spin chain.}

Now, as argued in \cite{JRS}, the 
 integrable model based on the four-dimensional fundamental 
representation of $osp(2/2)$ sits inside the Goldstone phase of this 
model, and goes to the supersphere sigma model $OSP(2/2)/OSP(1/2)$ in 
the scaling limit. The latter model can be described  \cite{WeheSal} by the  following 
action. Parameterize the $S^{1,2}$ supersphere as
\begin{eqnarray}
    \phi_{1} &=& \cos\phi\left(1-\eta_{1}\eta_{2}\right)\nonumber\\
    \phi_{2} &=& \sin\phi\left(1-\eta_{1}\eta_{2}\right)
\end{eqnarray}
such that $\phi_{1}^{2}+\phi_{2}^{2}+2\eta_{1}\eta_{2}=1$. The action is then
\begin{equation}
    S={1\over g}\int 
    {\rm d}^2x \, \left[\left(\partial_{\mu}\phi\right)^{2}(1-2\eta_{1}\eta_{2})+2\partial_{\mu}\eta_{1}\partial_{\mu}\eta_{2}
    -4\eta_{1}\eta_{2}\partial_{\mu}\eta_{1}\partial_{\mu}\eta_{2}\right]\label{sigmaact}
\end{equation}
where $\phi$ is compactified, $\phi\equiv \phi+2\pi$.  The integrable 
chain corresponds to 
spontaneously broken symmetry, with the theory being 
free in the UV and massless in the IR, corresponding to a {\sl 
negative} coupling constant  $g$. Critical properties are described 
by the $g\rightarrow 0$ limit of Eq.~(\ref{sigmaact}). A rescaling and 
relabeling brings Eq.~(\ref{sigmaact}) into the form
\begin{equation}
    S=\int 
    {\rm d}^2x \, \left[-\left(\partial_{\mu}\phi\right)^{2}(1+2|g|\eta_{1}\eta_{2})+2\partial_{\mu}\eta_{1}\partial_{\mu}\eta_{2}
    +4|g|\eta_{1}\eta_{2}\partial_{\mu}\eta_{1}\partial_{\mu}\eta_{2}\right]\label{sigmaacti}
\end{equation}
with now $\phi\equiv\phi +{2\pi\over\sqrt{|g|}}$. As $g\rightarrow 0$, we 
thus obtain a symplectic fermion and a non-compact boson, with the 
wrong sign of the metric. This is exactly the low-energy theory we 
had  identified earlier for the $Q=0$ antiferromagnetic Potts model, 
thus confirming our picture. 

Note that the presence of  a 
continuous spectrum in this $osp(2/2)$ version of the problem
has a physical origin in the fact 
that the symmetry is spontaneously broken, and thus correlation 
functions of order parameters have no algebraic decay (though  
have logarithmic behaviour). This feature was verified 
numerically in \cite{JRS} where it was found that the scaling 
dimensions of $p$ leg polymer operators  extracted through transfer 
matrix calculations vanish inside the entire 
Goldstone phase indeed. 

As a final note we remark that the Bethe equations for the $osp(2/2)$ 
chain have been most often \cite{MR,RM} written in a different grading from the 
one we used. It is also interesting to compare the analysis of the 
low energy physics from the more standard point of view.

The usual  Bethe equations are written as 
\begin{eqnarray}
    \left({\alpha-i\over
    \alpha+i}\right)^L&=&
    \prod {\alpha-\beta-2i\over\alpha-\beta+2i}\nonumber\\
    \prod {\beta-\alpha-2i\over\beta-\alpha+2i}&=&
    \prod {\beta-\beta'-4i\over\beta-\beta'+4i}
    \end{eqnarray}
   with the energy
 \begin{equation}
     E= {4\over\pi}\sum {1\over 1+\alpha^{2}}
 \end{equation}
(note the change of sign compared to the other Bethe ansatz). The relevant solutions for the low energy physics are: $\alpha$ two-strings centered on real $\beta$'s (denoted 
then by $y$), and  
real $\alpha$'s (which are not in complexes, denoted then by $z$). The 
corresponding  Bethe 
equations are
\begin{eqnarray}
   \left({y-3i\over y+3i}{y+i\over 
   y-i}\right)^{L}&=&
   \prod_{y'}{y-y'-4i\over 
   y-y'+4i}\prod_{z}{y-z-2i\over 
   y-z+2i}\nonumber\\
   \left({z-i\over z+i}\right)^{N}&=&
       \prod_{y}{z-y-2i\over z-y+2i}
\end{eqnarray}
The ground state is filled with the complexes. Going to the continuum limit and
using Fourier transform gives the following physical equations, if we 
call $\rho$ the density of complexes, $\sigma$ the density of real 
particles
\begin{eqnarray}
\hat{\rho}+\hat{\rho}^h&=&{1\over 2\cosh x}- {{\rm e}^{-|2x|}\over 2\sinh 
2|x|}\hat{\rho}^h
+{1\over 2\sinh 2|x|}\hat{\sigma}\nonumber\\
\hat{\sigma}+\hat{\sigma}^h&=&{1\over 2\cosh x} - {{\rm e}^{-2|x|}\over 2\sinh
2|x|}\hat{\sigma}
+ {1\over 2\sinh 2|x|}\hat{\rho}^h
\end{eqnarray}
These equations show that the chain has a relativistic limit (a 
feature peculiar to the value $b=0$ of the fundamental 
representation \cite{Saleurslnk}). Moreover, the density equations coincide with those for the 
$Q\rightarrow 0$ antiferromagnetic Potts model discussed above: even 
though we have started from two different  gradings, we end up with 
identical physical equations, a very strong indication that our 
approach is consistent \cite{Essler} (note in particular that
the quantum numbers of the 
excitations, being expressed naturally in terms of complexes and real 
holes, have different expressions in terms of the bare quantum 
numbers) and the results correct.

To finish the analysis of the $osp(2/2)$ system, 
the most useful quantity one can consider is the `equivalent spin' 
defined by forming the combination $m\equiv b-3S^{z}$. The four 
states of the fundamental representation $b=0, S^{z}=\pm 
1/2;S^{z}=0,b=\pm 1/2$ correspond respectively to the values 
$m=3/2,1/2,-1/2,-3/2$. On the other hand, according to the detailed 
lattice model derivation of the Bethe ansatz equations \cite{RM,MR}, 
$$
m=b-3S^{z}=\#\beta-\#\alpha-{L\over 2}
$$
The spectrum is expressed in terms of sums and differences of $\delta 
n_{c}^{h}$ and $\delta n_{r}$, where the subscripts $c,r$ indicate 
complexes and real particles respectively. One has
\begin{eqnarray}
   \delta n_{c}^{h}-\delta n_{r}&=&\delta n_{\beta}-\delta 
   n_{\alpha}\nonumber\\
   \delta n_{c}^{h}+\delta n_{r}&=&\delta n_{\beta}-3\delta n_{\alpha}
\end{eqnarray}
We thus see that it is possible to change the value of the 
pseudospin $m$ without changing $\delta n_{c}^{h}+\delta n_{r}$, i.e., 
by exciting only the flat part of the spectrum. In other words, in 
the geometrical interpretation,  the 
gaps for $p$ leg operators are all expected to vanish. This result 
agrees with predictions based on the Goldstone analysis of the 
$OSP(2/2)$ loop model, as well as numerical calculations \cite{JRS}, 
and we take the consistency of the different approaches as a 
justification of our use of the dressed charge formalism in these 
problems.

\subsection{Explicit mapping  between the two models}

Even though we have found  that the Bethe equations for the two models are 
identical, we do not exactly know how to map them onto each other. 
Clearly, one must combine pairs of neighbouring spin $1/2$ 
representations of the vertex model to make up four-dimensional 
representations of the $osp(2/2)$ algebra, but we are not sure how to 
implement this. Part of the difficulty comes from the highly singular 
nature of the Potts transfer matrix as $Q\rightarrow 0$. 

\begin{figure}
	 \centering
	     \includegraphics[width=2in]{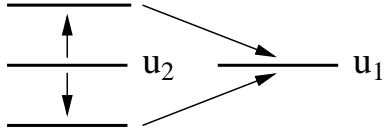}
	 \caption{Indecomposable representation $1/2^{\otimes 2}$ for $t=i$. }
	 \label{indecrep}
	 \end{figure}

It is interesting to notice however that as $t\rightarrow i$, the 
tensor product of  two fundamental representations of $U_{t}sl(2)$, 
which  decomposes as the sum of a one and a three dimensional 
representations like for $t=1$ when $t$ is generic, becomes 
indecomposable, and has the structure shown in Fig.~\ref{indecrep}, where 
$|u_{1}\rangle={\rm e}^{i\pi/4}|+-\rangle+{\rm e}^{-i\pi/4}|-+\rangle$ and 
$|u_{2}\rangle={\rm e}^{-i\pi/4}|+-\rangle+{\rm e}^{i\pi/4}|-+\rangle$, and the 
arrows indicate the action of the $U_{t}sl(2)$ generators:
\begin{eqnarray}
    S^{-} &=& t^{\sigma^{z}}\otimes \sigma^{-}+\sigma^{-}\otimes 
    t^{-\sigma^{z}}\nonumber\\
    S^{+} &=& t^{\sigma^{z}}\otimes \sigma^{+}+\sigma^{+}\otimes 
	t^{-\sigma^{z}}
	\end{eqnarray}

It turns out that, if one introduces the other generators
\begin{eqnarray}
    \tilde{S}^{-} &=& t^{-\sigma^{z}}\otimes \sigma^{-}+\sigma^{-}\otimes 
    t^{\sigma^{z}}\nonumber\\
    \tilde{S}^{+} &=& t^{-\sigma^{z}}\otimes \sigma^{+}+\sigma^{+}\otimes 
	t^{\sigma^{z}}
	\end{eqnarray}
at $t=i$, together with the limits $\lim_{t\rightarrow i}{(S^{\pm})^{2}\over 
t+t^{-1}}$, one obtains the four-dimensional representation of 
$osp(2/2)$. This suggests  a  mechanism by which the symmetry of the 
lattice model must be enhanced at the critical point. One can in fact 
show by brute force that by concatenating pairs of spins $1/2$ into 
four degrees of freedom and carefully taking the $Q\rightarrow 0$ 
limit of the Boltzmann weights of the heterogeneous 6-vertex model 
equivalent to the critical antiferromagnetic $Q$ state Potts model, 
the $R$ matrix of the $osp(2/2)$ integrable vertex model in the 
fundamental ($[b=0,j=1/2]$) representation is indeed obtained. This 
will be discussed elsewhere.

Note of course that the $osp(2/2)$ integrable model can be reformulated as a loop model. 
In fact the whole Goldstone phase has been argued to be realized 
physically by taking a model made of a single dense loop where 
crossings are allowed. This can be  obtained (and studied numerically) 
for instance by taking the usual \cite{Baxterbook} covering of the 
surrounding lattice of the square lattice with a single self avoiding 
loop, and then allowing self-crossings. 

\begin{figure}
\begin{center}
 \leavevmode
 \epsfysize=70mm{\epsffile{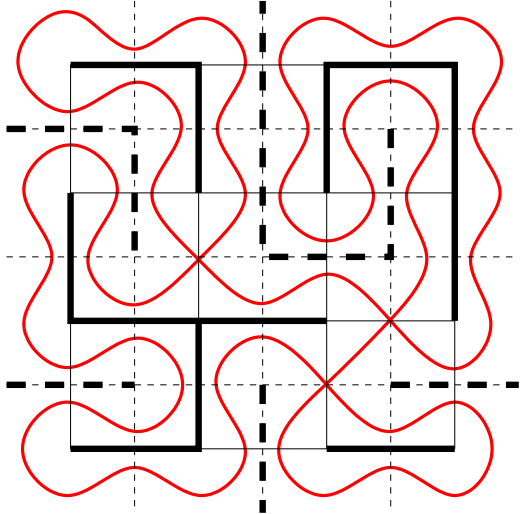}}
 \end{center}
 \protect\caption[3]{Mapping between forests and self-intersecting loops.
Thin solid (resp.\ dashed) lines represent the lattice ${\cal L}$
(resp.\ the dual lattice ${\cal L}^D$). All edges of ${\cal L}^D$ penetrating
the boundary of ${\cal L}$ are understood to join at a unique dual site
situated on the infinite, exterior face of ${\cal L}$. Thick solid lines
depict the forest (here with four component trees); thick dashed lines
depict the (unique) dual tree. The grey (red online) solid line is a
self-intersecting loop on the surrounding lattice (see text). It bounces
off occupied edges (thick linestyle) on ${\cal L}$ and on ${\cal L}^D$, and
self-intersects whenever a pair of intersecting direct and dual edges
are both non-occupied.} 
\label{figloop}
\end{figure}

It is well known that in the case where self-crossings are not 
allowed, the loop the hull of a spanning tree, hence giving rise to 
the well-known close relationship between the critical theory of 
dense polymers and that of spanning trees, both being related to 
symplectic fermions \cite{DuplSal,DavDupl}.  This relationship can to some extent be 
generalized to the arboreal gas and the loop with self-crossings (see Fig.~\ref{figloop}). 
Indeed, draw on  the lattice ${\cal L}$ a spanning forest. Draw on 
the dual lattice ${\cal L}^{D}$ (which in particular contains a single site 
outside ${\cal L}$) a spanning tree. The resulting contour on the surrounding 
lattice is a single loop with a number of self-intersections equal to 
the number of trees on ${\cal L}$ plus one! So we have a sort of 
correspondence between a self-crossing loop and an arboreal gas, i.e.,
between a system with $osp(2/2)$ symmetry and the $Q\rightarrow 0$ 
antiferromagnetic Potts model. However,  
this is not a very satisfactory correspondence. In particular, because the tree 
on the dual lattice plays a crucial role in determining the loop 
configuration, there is no one to one 
correspondence between loop and forest. Also, giving a negative fugacity to 
trees corresponds to a negative weight for self-crossings of the loop, 
while the existence of the Goldstone phase was argued in \cite{JRS} 
for positive weights only.

\section{Conclusion}

The most interesting conclusion is probably the emergence of a non-compact
boson in the continuum limit of the critical antiferromagnetic 
Potts model at $Q=0$. Although it is difficult to demonstrate the 
existence of such a non-compact object based on any single analysis, 
the convergence  of Bethe ansatz calculations, numerical 
calculations, and symmetry considerations, leaves very little doubt 
that the conclusion is correct. This is, to our knowledge
(together with our earlier work on Goldstone phases \cite{JRS}, which as we saw 
is intimately related), the first 
time that a non-compact object appears in the continuum limit of a 
lattice model with  a finite number of degrees of freedom (but
`unphysical' Boltzmann 
weights). In a subsequent work, we will see that this feature extends 
to the 
antiferromagnetic Potts model away from $Q=0$ \cite{bigantiferro}. 

We have also seen how this non-compact boson is part of a free 
$OSP(1/2)$ theory, which appears here as a critical point of the 
super (hemi-)sphere sigma model. The main question that remains 
unanswered in this study of  the physics of models with 
supergroup symmetry is 
what happens to the `other' (Nienhuis) critical $O(n=-1)$ 
model \cite{Nienhuis}, and how this is related with a  possible difference between the critical points in the 
hemi and full supersphere sigma models. We hope to get back to these 
questions soon.

To finish, we can clarify several remaining points. The first is the 
issue of what happens when the size of the lattice system $N$ is odd. As will 
be discussed in details for the general antiferromagnetic Potts model 
in \cite{bigantiferro}, the spectrum in that case does not show any 
signs of non-compactness because the bosonic field $\phi$ appears to 
be  twisted (and thus its properties independent of any 
compactification radius). By $OSP(1/2)$ symmetry, the fermionic field 
is also twisted, so the effective central charge in this sector is 
\begin{equation}
 c_{\rm eff}=c-24 h_{\rm tw}^{\phi}-24 h_{\rm tw}^{\eta}=
 -1-\frac32+3=\frac12
    \end{equation}
in agreement with numerical calculations \cite{JSS,bigantiferro}.

The second point is that 
numerical \cite{JSS} as well as analytical arguments  
\cite{bigantiferro} suggest that the critical point is, in the Potts 
model point of view,  
approached from the Berker-Kadanoff phase (the broken symmetry 
phase) with a singularity in the free energy of the form
\begin{equation}
    f\propto (a^{*}-a)\ln(a^{*}-a) \quad \mbox{as \ }
    a\uparrow a^{*}.\label{freeenergy}
    \end{equation}
(We recall that $a^*=-4$ for the square lattice.)
Using hyperscaling, this corresponds to a perturbation of the  
critical point by a field of {\sl vanishing dimension}. There are of 
course plenty of such fields in the non-compact boson theory. We 
speculate that the scaling limit near and below  $a^{*}$ could be 
described by the simple Landau-Ginzburg theory 
\begin{equation}
    S=\int d^{2}x 
    \left[(\partial_{\mu}\phi)^{2}+2\partial_{\mu}\eta_{1}\partial_{\mu}\eta_{2}\right]+
    (a^{*}-a)\left(\phi^{2}+2\eta_{1}\eta_{2}-1\right)
    \end{equation}
 for which the logarithm in $f$ [see Eq.~(\ref{freeenergy})] follows as well. 

\begin{table}
 \begin{center}
 \begin{tabular}{lr|rrrrr}
         & $a$&  $L=6$  &  $L=8$  &  $L=10$ &  $L=12$ & $L=14$ \\ \hline
  $x_1$: &  0 &  0.0000 &  0.0000 &  0.0000 &  0.0000 & 0.0000 \\
         & -1 & -0.0859 & -0.0792 & -0.0747 & -0.0714 & \\
         & -2 & -0.1092 & -0.0992 & -0.0925 & -0.0877 & \\
         & -3 & -0.1168 & -0.1079 & -0.1003 & -0.0947 & \\
         & -4 &  0.0000 &  0.0000 &  0.0000 &  0.0000 & 0.0000 \\ \hline
  $x_2$: &  0 &  0.7492 &  0.7498 &  0.7499 &  0.7500 & \\
         & -1 &  0.5823 &  0.5975 &  0.6075 &  0.6146 & \\
         & -2 &  0.5216 &  0.5492 &  0.5663 &  0.5781 & \\
         & -3 &  0.4936 &  0.5253 &  0.5464 &  0.5609 & \\
         & -4 &  0.5466 &  0.5691 &  0.5830 &  0.5927 & \\ \hline
  $x_3$: &  0 &  1.9671 &  1.9894 &  1.9958 &  1.9980 & 1.9989 \\
         & -1 &         &  1.6959 &  1.7213 &  1.7384 & \\
         & -2 &  1.4930 &  1.5727 &  1.6201 &  1.6514 & \\
         & -3 &  1.4050 &  1.5062 &  1.5667 &  1.6065 & \\
         & -4 &  1.5268 &  1.6157 &  1.6621 &  1.6907 & 1.7105 \\ \hline
 \end{tabular}
 \caption{Scaling exponents $x_k=2h_k$ describing the probability
 $P(r) \sim r^{-2x_k}$ that $k$ distinct trees come close in points
 $0$ and $r$. The data is obtained by fitting transfer matrix results
 in the random-cluster picture, on periodic square-lattice cylinders
 of widths $L$ and $L-2$. Note the strong finite-size dependence for
 tree fugacity $a \neq 0$. For $a=0$ the exact result is
 $x_k = (k^2-1)/4$ \cite{DuplSal}.}
 \end{center}
\end{table}

The last point concerns the physical interpretation of our results. 
This of course has to be taken with a bit of care, as our model involves 
negative Boltzmann weights. But the image we have is that in the 
region $a^*<a\leq 0$, the partition function is dominated by 
configurations where a single infinite tree covers a finite fraction
of the lattice---this fraction is unity as $a\rightarrow 0$, and 
remains finite in the whole phase $a^*<a<0$, the rest of the lattice 
being covered by trees of finite size. Tree exponents are the same as 
in the limit $a\rightarrow 0$, as can be checked numerically in
Table~1. In particular, the fact that the one-tree exponent remains 
equal to zero guarantees that there is an infinite tree, and that 
its fractal dimension is equal to two. At $a=a^*$, we have coexistence 
of two phases, one with an infinite tree of finite density, the other 
with only finite trees. As $a=a^*$ is crossed,  the 
derivative of the free energy with respect to $a$ is discontinuous \cite{JSS}, implying that 
the average number of trees per lattice site, and thus the average 
size of finite trees, experiences a discontinuity. In the phase 
$a<a^*$ (like in the phase $a>0$), there are only finite trees, of 
typical linear size given by the finite correlation length. The 
point $a=a^*$ is actually a first-order critical point, that is a 
point where a first and second order phase transition  coincide 
\cite{JSS}. At such a point, some correlations still decay 
algebraically, while some of the thermodynamic properties have 
discontinuities. The latter translate into operators of vanishing 
dimensions, for which we expect an infinite multiplicity. The 
simplest formalism to encompass this multiplicity is a non-compact 
boson, in agreement with what was  found previously. 

We conclude with a side remark. The fermionic 
representation of the arboreal gas model, instead of being used to 
argue for the presence of an $OSP(1/2)$ symmetry,  can be used to 
get yet another representation, this time in terms of a loop model. 
For this, go back to the initial path integral (\ref{identity}) written
in terms of fermions $\eta_{1},\eta_{2}$ only 
\begin{equation}
   \exp\left[(4+t)\sum_{i}\eta_{1}(i)\eta_{2}(i)-\sum_{\langle ij \rangle}(\eta_{1}(i)\eta_{2}(j)+
   \eta_{1}(j)\eta_{2}(i))-t\sum_{\langle ij \rangle}
   \eta_{1}(i)\eta_{2}(i)\eta_{1}(j)\eta_{2}(j)\right]
\end{equation}
Simply expand the exponent, and contract the fermions. 
This is like the standard high temperature expansion 
for the Nienhuis $O(n)$  model . It gives rise to a sum over 
non-intersecting loops, with 
weight $-2$ per loop (like in the $O(n=-2)$ model),
plus non-overlapping monomers with weight $4+a$, 
and dimers  (which can be considered as degenerate length-two loops) with weight $-a$. The critical point $a=-4$ means monomers 
disappear, dimers have weight $4$. We are not sure why this model 
should be critical---note the dimers seem to encourage swelling of the 
loops, instead of self attraction as one would expect to get to a 
tricritical point---and this direction certainly deserves further 
study. Note that a critical point with $c=-1$ has been identified in 
a modified version of the $O(n=-2)$ model in Ref.~\cite{Blote}. \\



\noindent {\bf Acknowledgments:} This work was supported by the 
Department of Energy 
(HS). We thank F. David and A. Sokal for discussions.

\begin{appendix}
    
    \section{Bethe equations}
    
    The standard integrable model based on  the $sl(2/1)=osp(2/2)$ 
    superalgebra is the 
    Perk Schultz model. It uses 
    the fundamental representation of dimension 3, and leads to the Bethe ansatz   equations (for a review see \cite{EK})
    \begin{eqnarray}
    \left({\alpha-2i\over
    \alpha+2i}\right)^N=
    \prod {\alpha-\beta-2i\over \alpha-\beta+2i}\nonumber\\
    \prod {\beta-\alpha-2i\over \beta-\alpha+2i}=
    \prod {\beta-\beta'-4i\over \beta-\beta'+4i}\label{bethei}
    \end{eqnarray}
    Algebraically, these equations correspond to treating the fundamental 
    representation of $sl(2/1)$ as a non-typical 
    fermionic representation  with parameters $[b,j]=[-1/2,1/2]$, 
    of dimension 3 instead of  4. One could as well consider a model 
    based on higher spin representations $[b,j]=[-j,j]$ obtained by 
    taking the fully supersymmetric component in the product of $n=2j$ 
    fundamental representations. The equations would read as 
    Eq.~(\ref{bethei}) but on the left-hand side the factor $2$ would be 
    replaced by $2n$. 
    
    More generally, for the four-dimensional typical representations 
      $[b,j=1/2]$,  
      the Bethe equations generalizing Eq.~(\ref{bethei}) read 
      \cite{Ziad,PfF,BGLZ}
      \begin{eqnarray}
      \left({\alpha+i(2b-1)\over
      \alpha-i(2b-1)}\right)^N=
      \prod {\alpha-\beta-2i\over\alpha-\beta+2i}\nonumber\\
      \prod {\beta-\alpha-2i\over\beta-\alpha+2i}=
      \prod {\beta-\beta'-4i\over\beta-\beta'+4i}
      \end{eqnarray}
 In particular, the case $b=0$ is similar to Eq.~(\ref{bethei}) but on 
 the left-hand side the factor $2$ is replaced by $1$ corresponding 
 roughly to taking a `quarter spin'.

      The second type of Bethe ansatz for the Perk Schultz model is 
    obtained when associating the fundamental  representation to the bosonic root of 
    the distinguished Dynkin diagram instead. One obtains then 
the so called Sutherland's equations
    \begin{eqnarray}
    1=
    \prod {\alpha-\beta-2i\over \alpha-\beta+2i}\nonumber\\
    \left({\beta-2i\over
    \beta+2i}\right)^N\prod {\beta-\alpha-2i\over \beta-\alpha+2i}=
    \prod {\beta-\beta'-4i\over \beta-\beta'+4i}\label{betheii}
    \end{eqnarray}
Finally, it is also possible to write a Bethe ansatz based on the alternative 
    Dynkin diagram, where both roots are fermionic
    \begin{eqnarray}
    1=
    \prod {\alpha-\beta-2i\over\alpha-\beta+2i}\nonumber\\
    \left({\beta-2i\over
    \beta+2i}\right)^N=
    \prod {\beta-\alpha-2i\over\beta-\alpha+2i}\label{betheiii}
    \end{eqnarray}
    In this ansatz, equations for a mix of the fundamental 
    and its conjugate are particularly elegant \cite{Links}:
    \begin{eqnarray}
    \left({\alpha-2i\over
    \alpha+2i}\right)^N=
    \prod {\alpha-\beta-2i\over \alpha-\beta+2i}\nonumber\\
    \left({\beta-2i\over
    \beta+2i}\right)^N=
    \prod {\beta-\alpha-2i\over \beta-\alpha+2i}
    \end{eqnarray}
As for the model based on the four-dimensional representation, its 
equations are given in the text as Eq.~(\ref{frahmbethe}) and correspond once again to taking 
a `quarter spin'.


\end{appendix}

\end{document}